\documentclass[aps,twocolumn,epsf,floats,prl,reprint,nofootinbib]{revtex4-1}
\usepackage{graphics,graphicx,epsfig}
\usepackage{amssymb,color}
\usepackage{epsf,epstopdf,wrapfig}
\usepackage{amsmath}
\usepackage{multirow}

\usepackage{color,graphicx}

\usepackage{amsmath}

\begin{document}

 \title{Critical fluctuations in proteins native states}
 
\author{Qian-Yuan Tang$^{1}$} 
\author{Yang-Yang Zhang$^{1}$}
\author{Jun Wang$^{1}$} 
\author{Wei Wang$^{1}$}
\author{Dante R Chialvo$^2$} 


\affiliation{$^1$ National Laboratory of Solid State Microstructure, Department of Physics, and Collaborative Innovation Center of Advanced Microstructures, Nanjing University, Nanjing 210093, China}
\affiliation{$^2$ Consejo Nacional de Investigaciones Cient\'ificas y Tecnol\'ogicas (CONICET), Godoy Cruz 2290, Buenos Aires, Argentina}  

\date{\today}

\begin{abstract}
We study a large data set of protein structure ensembles of very diverse sizes determined by nuclear magnetic resonance.  By examining the distance-dependent correlations in the displacement of residues pairs and conducting finite size scaling analysis it was found that  the correlations and susceptibility behave as in systems near a critical point implying that,  at the native state,  the motion of each amino acid residue is felt by every other residue up to the size of the protein molecule. Furthermore certain protein's shapes corresponding to maximum susceptibility were found to be more probable than others.  Overall the results suggest that the protein's native state is critical, implying that despite being posed near the minimum of the energy landscape, they still preserve their dynamic flexibility. 
\end{abstract}
 
\maketitle

Protein molecules are formed by large unbranched chains of amino acids, which turn into a complex folded shape as free energy is minimized.
It is this highly specific three-dimensional folded structure, known as native state, that makes the protein capable of performing its biological function \cite{review1}.  Proteins carry out their functions by switching from one shape to another, even transiently, as for instance when it recognizes and binds with another molecule. To achieve such performance the structure of the native state must be very susceptible to sense the signal and switch to another shape, but also be stable enough to warrant reproducibility. It is well known that these apparently contradictory demands are exhibited by systems near a critical point because of the coexistence of maximum susceptibility and  long range correlations \cite{bak, mora, reviewBiophys,chialvo2010,chate}.
 
These views are discussed on a number of recent reports emphasizing different aspects of critical fluctuations in the protein equilibrium dynamics. This includes the geometric properties \cite{moret}, the slowness in relaxation in the dynamics of large biomolecules \cite{lu}, the role of their low-frequency global modes \cite{bahar1998, bahar2010} in the proteins' functional dynamics,  the overlap of the large-scale conformational change in allosteric transitions and the low frequency normal modes\cite{yang}, the role of the water surrounding the molecule \cite{chalmers}, as well as the near-critical states emerging in the sequential correlations of protein families \cite{mora}.
 
Although it is often recognized that the available data seems still far from being the ideal to test for criticality, we propose here an approach to investigate this issue. We use a large number of protein structure ensembles determined by solution nuclear magnetic resonance (NMR).  Since each ensemble contains different structures of the same protein, the basic idea is to assume that each of structures can be seen as a hypothetical instantiation of the spontaneous conformational changes that the protein exhibit through time. By examining the distance-dependent correlations in the displacement of residue pairs and conducting  finite size scaling analysis it is shown that  the correlations and susceptibility behave as expected in systems near a critical point. The results imply that at the native state,  the motion of each and every amino acid residue is felt by every other residue, up to the size of the protein molecule.  

\begin{figure}[htbp]
\begin{center}
\includegraphics[width=.495 \textwidth]{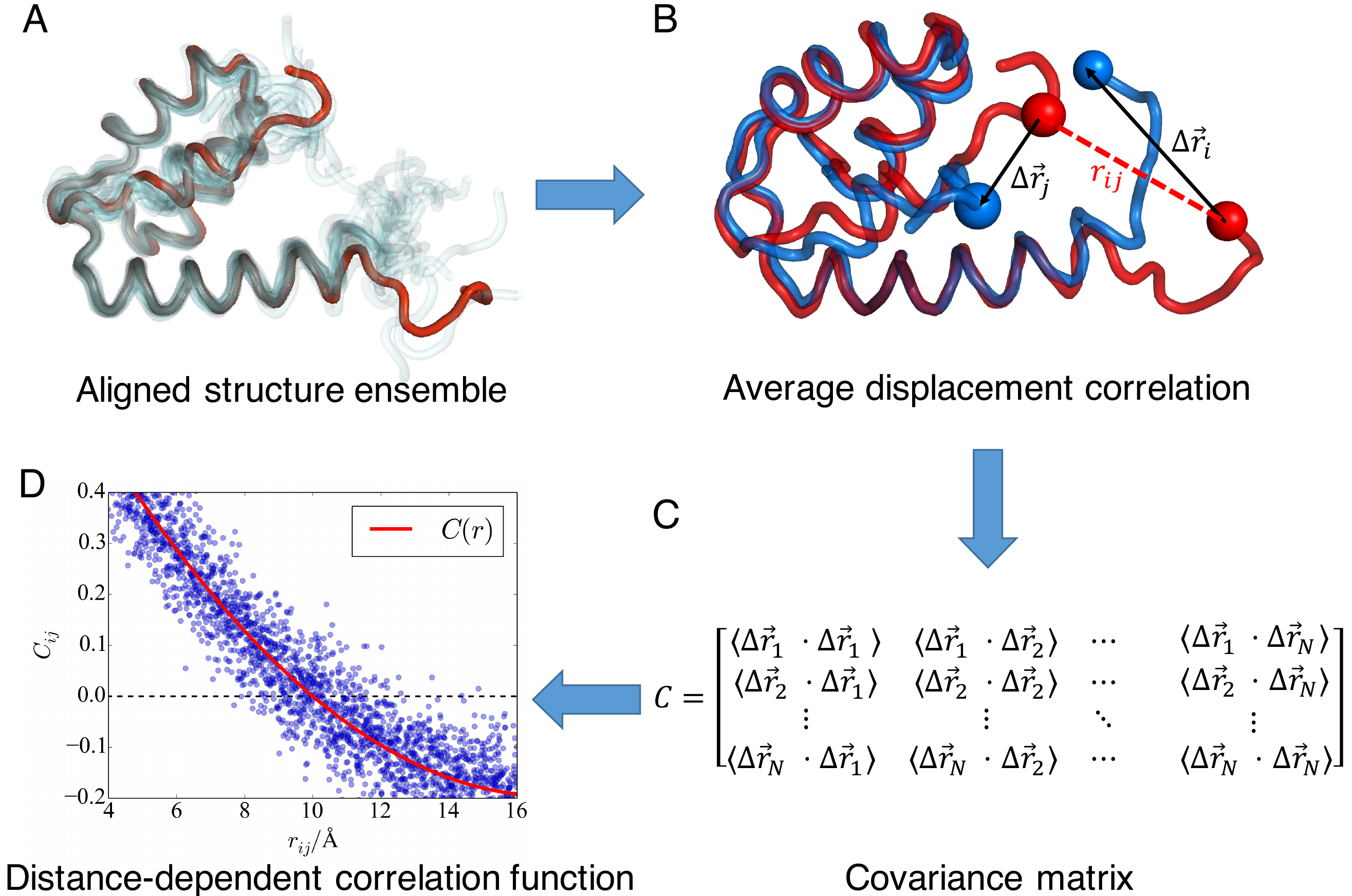}
\caption{(A) Aligned C$_\alpha$ traces of the structure ensemble of a protein molecule (PDB code: 2KQ6) determined by NMR. 
(B) An illustration of the definition of average displacement correlation $C_{ij} = \langle \Delta \vec{r}_i \cdot \Delta \vec{r}_j \rangle$. (C) The elements in a covariance matrix $C$. (D) The scattering plot of the $r_{ij}$ and $C_{ij}$ and the averaged distance-dependent correlation function $C(r).$ }
\label{cartoon}
\end{center}
\end{figure}

\emph{Fluctuations and correlations: Data and definitions.} The dataset analyzed contains 7678 protein structure ensembles with not less than 10 different structures from the protein data bank (PDB)\cite{PDB} (see complete list and details in the Supp. Info.). For each structural ensemble, we selected one configuration as a reference state, then by doing 3D structure alignment, the degrees of freedoms related to the translational and rotational motion are removed. As Fig.\ref{cartoon}(A) shows one conformation (colored in red) is set as the reference state, and the other conformations were aligned to that reference state. After the alignment, the displacement of every atom from the reference state was computed (Fig.\ref{cartoon}(B)). The calculations are based on the python package ``ProDy'' \cite{bakan}.

For simplification, we mainly focus on the C$_\alpha$ traces of the proteins. For a protein molecule made up of $N$ amino acid residues, for all residue pairs $i$ and $j$ ($1\leq i, j\leq N$), the distance $r_{ij}$ between the two residues is approximated as the distance between the two C$_{\alpha}$'s, and the average correlation $C_{ij}$ (the elements of the covariance matrix) of the displacement of the two residues $i$ and $j$ is approximated by the average inner product of the displacement of two C$_\alpha$ atoms in the two residues. 

In this manner it is constructed the covariance matrix (Fig.\ref{cartoon}(C)), $C_{ij} = \langle \Delta \vec{r}_i \cdot \Delta \vec{r}_j \rangle$, where $\Delta \vec{r}_i $ is the displacement from the average configuration of the C$_\alpha$ atom in residue $i$ and $\Delta \vec{r}_j $ is the similar displacement for atom $j$.  From the covariance $C_{ij}$, the orientational correlation of residue pairs $\phi_{ij} = {C_{ij}}/{\sqrt{C_{ii} \cdot C_{jj}}}$ is obtained. Here, $C_{ii}$ and $C_{jj}$ are the auto-correlation of the displacement of residue $i$ and $j$, which are proportional to the B factors (root mean square fluctuations) of the C$_\alpha$ atoms. 

For our protein data,  $C_{ij}$ (and $\phi_{ij}$) is roughly a function $C(r)$ of the distance $r_{ij}$ (as shown in Fig.\ref{cartoon}(D)). Thus, for any given protein molecule, one calculate the average covariance and crosscorrelation for all the residue pairs that have similar distance between two C$_\alpha$ atoms ($r_{ij}\approx r$), allowing to define the distance-dependent covariance $C(r)$ and crosscorrelation $\phi(r)$ of single protein molecules as:
\begin{equation}
C(r) = \frac{\sum_{i,j}C_{ij} \delta(r-r_{ij})}{\sum_{i,j} \delta(r-r_{ij})}; \\
\phi(r) = \frac{\sum_{i,j}\phi_{ij} \delta(r-r_{ij})}{\sum_{i,j} \delta(r-r_{ij})}.
\end{equation}
Moreover, to reveal the general properties in the dynamics of proteins with same sizes, the average distance-dependent correlation functions $C(r)$ and $\phi(r)$ could also be calculated by similarly averaging the $C_{ij}$ and $\phi_{ij}$ pairs for the proteins with similar radius of gyration $R_g$.

\begin{figure*}[t!]
\includegraphics[width=1.57\columnwidth]{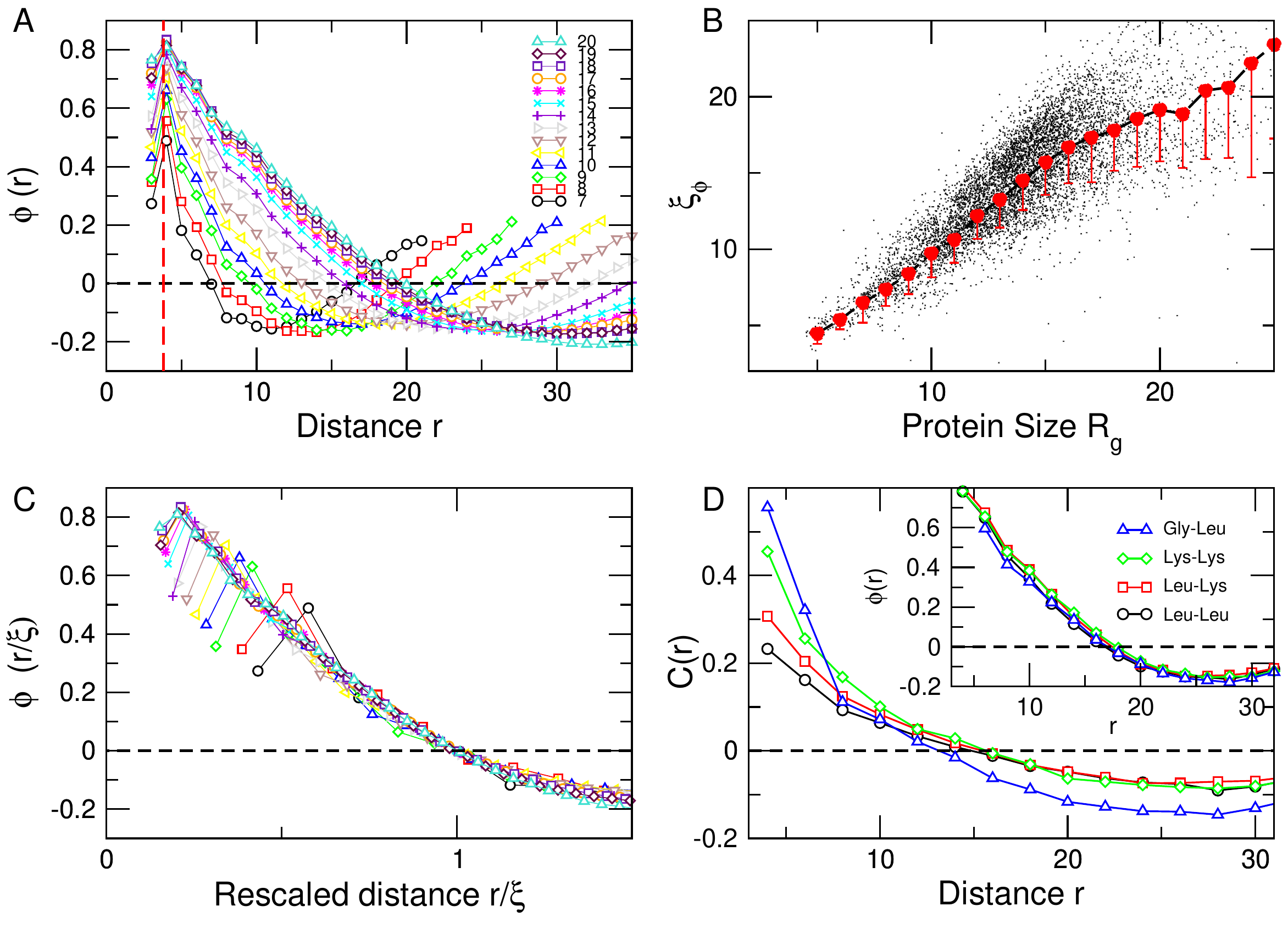}
\caption{(A) The distance-dependent crosscorrelation function $\phi(r)$ averaged for proteins with similar $R_g$. 
(B) Scattering plot of correlation length $\xi_\phi$ as a function of the average $R_g$, where the red symbols  show the average $\xi= f(R_g)$ with the error bars denoting the standard deviation. 
(C) The scaling plot of $\phi(r/\xi)$. 
(D) The average distance-dependent covariance $C(r)$ and crosscorrelation function $\phi(r)$ (inset), for specific kinds of residues in proteins of $R_g\approx 15\AA$.}
\label{correlation}
\end{figure*}
\emph{Correlation length is proportional to protein size:} As shown in Fig.\ref{correlation}, for proteins with different $R_g$, the average crosscorrelation functions $\phi(r)$  exhibit a maximum (vertical dashed line in Fig.\ref{correlation}(A) at $r=3.8\AA$) which corresponds to the covalent bonds. Then, for  distances between residues close to $R_g$, the correlation function crosses zero,  which defines the correlation length $ \xi_\phi$ of $\phi(r)$ (Fig.\ref{correlation}(A)). (Since we focus on the crosscorrelation function, we just denote  $\xi_\phi$  as $\xi$). Beyond such a distance, the average correlation function is not vanishing, but first decreases to a negative minimum and then eventually approaches zero again. We note that this behavior of the correlation function is quite robust across different proteins;  it only differs at the  very long lengths due to boundary effects where the specific shape of each protein dominates the behavior of the correlation function.  This behavior means that within a protein (independently of its size), there is either strong correlation (short distance) or strong anticorrelation (large distance), but there is no region in which the correlation is consistently negligible. For all the proteins studied the correlation length $\xi_{\phi}$ and $R_g$ are found to be approximately proportional, (see Fig.\ref{correlation}(B)) which indicates that the residues of a protein are correlated to every other, independently of  how large the protein is. It means that if we could perturb a single residue, the consequence of such perturbation could be sensed by the entire protein molecule. 

Consequently, all the correlation functions computed at various sizes, can be rescaled by its gyration radius $R_g$ (or alternatively by its correlation length $\xi$). As shown in Fig.\ref{correlation}(C), all the curves collapse together after rescaling the distance between residues as $r/\xi$. 
Moreover, the distance-dependent covariance function $C(r)$ and the fluctuation of B-factors  also can be successfully rescaled, which indicates that not only the orientational crosscorrelation but also the amplitude of the fluctuations exhibits such kind of scale-free behavior (as occurs with the velocity correlations in the case of birds'  flocks described in Ref.\cite{cavagna2010}).

A more careful analysis reveals that the covariance correlations contain additional information about  residue-residue interactions.  For proteins with different sizes, different kinds of amino acid pairs would have different distance-dependent covariances, for example, as shown in Fig.\ref{correlation}(D), for leucine-leucine pairs, the covariance would in average be smaller than that of other listed types of residue pairs. This is  because, usually, hydrophobic residues are buried in the core of proteins so that the fluctuations would be small; and the zero points of the covariance are also slightly influenced by the type of amino acids, since glycine is a very small amino acid, the correlation length would be slightly deviated from the average. To refine the force field for coarse-grained models, one should take into consideration all the detailed residue specific interaction information, which is reflected in the distance-dependent covariance $C(r)$. However, for all types of residues pairs, the distance-dependent crosscorrelation $\phi(r)$ (as shown in Fig.\ref{correlation}(D) inset) still keeps the scale-free correlation with a similar correlation length.

\emph{Additional hints from finite size scaling:}
The type of correlations described above resembles those observed in the collective behavior of a variety of biological systems \cite{bak,mora,cavagna2010,attanasi}, in  which correlations are amplified by the vicinity to some critical point in the parameters space. However, most often,  the system size is very small respect to the thermodynamic limit, such that the value of the control parameter at which correlation and susceptibility peak depends on size. Thus in order to stay critical some inverse relation need to be found between a (pseudo) control parameter and the system size.  In turn, as the elegant demonstration of Attanasi et al.\cite{attanasi} shows, this finite-size scaling effect can be used to probe near-criticality.

Thus, we test such possibility for proteins of different sizes, performing  similar finite-size scaling as was done in Ref. \cite{attanasi}. To start, for each protein, the susceptibility is defined as the summed crosscorrelation between residue pairs, that is:
\begin{equation}
\chi = \frac{1}{N} \sum_{i\neq j}^N \phi_{ij} \cdot \theta(\xi_\phi - r_{ij}).
\end{equation}

Subsequently a dimensionless shape factor $s$ is defined as the pseudo control parameter of the protein, $s = Na^3/(L_aL_bL_c)$, where $a=3.8\AA$ is the size of a residue and $L_a$, $L_b$ and $L_c$ are the lengths of the principle axis of the protein ($L_a \leq L_b \leq L_c$). Such a parameter can also be understood as ``packing density'' because $L_aL_bL_c$ is proportional to the volume of an ellipsoid. For sphere-like protein molecules, the value of $s$ is relatively large (densely packed, and solid-like), while for elongated chains (loosely packed, and polymer-like), $L_c=Na$, and $L_a=L_b = a$, thus  $s = 1.$ 

\begin{figure} 
\includegraphics[width=.495\textwidth]{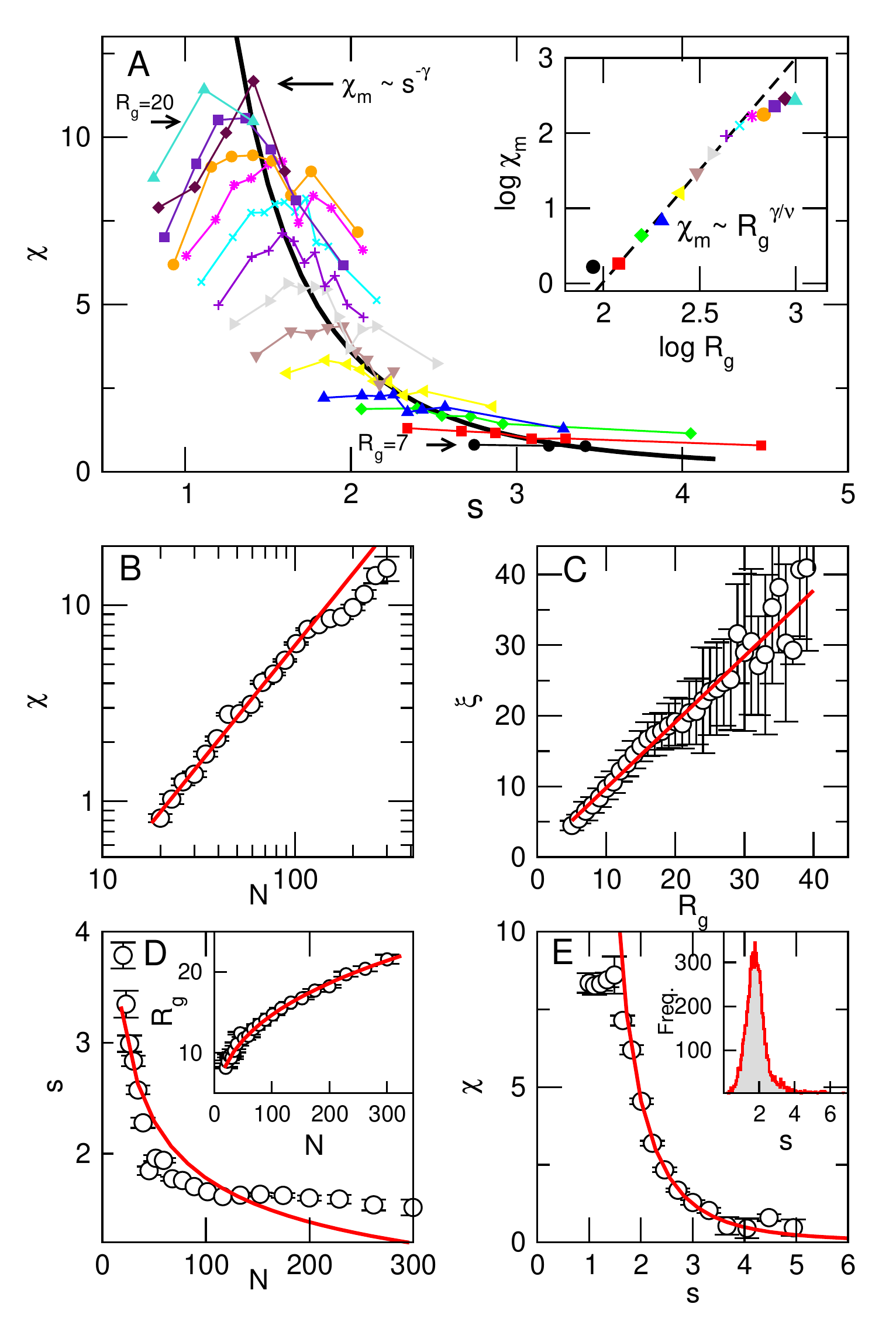}
\caption{Finite size scaling. (A) Susceptibility $\chi$ for proteins with different $R_g$ as a function of the control parameter, $s$. Each curve is calculated for proteins of similar $R_g$.  The peak of each proteins' susceptibility $\chi_m$ scales both with its shape (dark thick line) and size (inset). Same symbols and colors as in Fig. 2 indicating protein  size, $R_g.$
(B) Susceptibility, $\chi$, as a function of number of residues, $N$. (C) Correlation length, $\xi$, as a function of protein size, $R_g$. (D) Control parameter $s$ as a function of $N$. Inset shows $R_g$ as a function of $N$. (E) Susceptibility, $\chi$, as a function of $s$ in main plot, and the distribution of $s$ for all proteins in the inset.}
\label{Susceptibility}
\end{figure}

As shown in Fig.\ref{Susceptibility}(A), the computation of the susceptibility $\chi$ for proteins of similar $R_g$ reveals that  the $\chi-s$ plot exhibits a series of maximum $\chi_m$. Notice that when $R_g$ increases the shape factor $s$ for $\chi_m$  decreases, i.e., larger ``critical'' proteins seem to be more non-spherical than small ones. Also notice (in the inset) that susceptibility scales with protein size $\chi_m \sim R_g^{\gamma/\nu}$.

If the results correspond to (near) critical behavior then the following relations are expected to hold:
$s \sim N^{-1/3\nu}$, and $\xi \sim R_g$, as well as $\chi \sim  N^{\gamma/3\nu}$. Despite relatively large fluctuations the data exhibit  scaling behavior as shown in the fittings of Fig.3(B-D). For $s \sim R_g^{-1/\nu} \sim N^{-\alpha/\nu}$, and $R_g \sim N^{\alpha}$ we get $\alpha = 0.34$ (Panel B), thus $1/\nu \approx 0.96 \approx 1.09,$ $\nu \approx 1.04 \approx 0.9.$ Panel E  shows that for the relation $\chi \sim s^{-\gamma}$, $\gamma = 3.2$. Since  $\chi \sim R_g^{\gamma/\nu} \sim N^{\alpha\gamma/\nu}$, leads in both cases $\nu \approx 1.1$ (Panel D). Taking integers we could consider that $\alpha = 1/3$, $\nu = 1$, $\gamma = 3.$
 
Interestingly, it seems as if nature favors certain ``critical'' proteins, such that large and small proteins end up ``adjusting'' their shape in the folding process such that  they remain susceptible. The inset of Fig.\ref{Susceptibility}(E) shows that  the most frequent shape factor corresponds approximately to  the maximum susceptibility values.  A more detailed analysis shows (see the Supp. Info) that for proteins of similar $R_g$ value the most frequent $s$  corresponds to the maximum $\chi$ (i.e., at the critical line) at that $R_g$.

The scale free nature  of the correlations makes the distant residues well correlated with each other so that the global modes of the proteins could be easily excited.  That is to say that even though the shape of the proteins varies widely from one to another, it seems that the folding process towards the native state builds a shape for the proteins which ensures the emergence of long-range correlations, which is needed to produce conformational changes as well as to keep memory of the current configuration. 
 
Summarizing, the results show that the correlation length of the native state fluctuations is proportional to the gyration radius of the molecule, implying that the motion of any amino acid could influence all the others, up to the entire protein molecule. These results suggest that the proteins native states are not only posed near the minimum of the energy landscape, but also once there, they preserve the dynamic flexibility. In addition it is found that certain shapes are more probable, such that for any given protein size the folding process  favors  the shape with the maximum susceptibility (i.e.,  critical).

This work was supported by Natural Science Foundation of China (Grants 11334004, 11174133, 81421091) and National Basic Research Program of China (Grant No. 2013CB834100) and by CONICET of Argentina. Q-YT \& DRC acknowledge the hospitality of the Max Planck Institute for the Physics of Complex Systems at Dresden (Germany). Email addresses: QYT: tangqianyuan@gmail.com; JW: wangj@nju.edu.cn; WW: wangwei@nju.edu.cn; DRC: dchialvo@conicet.gov.ar.


\begin{thebibliography}{50}

 \bibitem{review1}J.R. Banavar and A. Maritan, \emph{Annu. Rev. Biophys. Biomol. Struct.} {\bf 36} 261--80 (2007).
\bibitem{bak}P. Bak, How nature works: the science of self-organized criticality (Springer, 2013).
\bibitem{mora}T. Mora and W. Bialek, \emph{J. of Stat. Phys.}  {\bf 144}, 268 (2011).
\bibitem{reviewBiophys} A.R. Honerkamp-Smith, S.L. Veatch, S.L. Keller, \emph{Biochimica et Biophysica Acta} {\bf 1788}, 53--63 (2009).
\bibitem{chialvo2010} D.R. Chialvo, \emph{Nature Physics} {\bf 6}, 744 (2010).
\bibitem{chate} H. Chat\'e and M. Mu\~noz, \emph{Physics} {\bf 7}, 120 (2014).
\bibitem{moret}M.A. Moret, \emph{Physica A} {\bf 390}, 3055--59 (2011).
\bibitem{lu}H.P. Lu, L. Xun,  X. S. Xie, \emph{Science} {\bf 282}, 1877 (1998).
\bibitem{bahar1998} I. Bahar, A.R. Atilgan, M.C. Demirel,  B. Erman, \emph{Phys. Rev. Lett.} {\bf 80}, 2733 (1998).
\bibitem{bahar2010} I. Bahar, T.R. Lezon, L.-W. Yang,  E. Eyal,\emph{Ann. Rev. of Biophysics} {\bf 39}, 23 (2010).
\bibitem{yang} L. Yang, G. Song,  R.L. Jernigan, \emph{Biophys. J.} {\bf 93}, 920 (2007).
\bibitem{chalmers} A.J. Patel, P. Varilly, S.N. Jamadagni, M.F. Hagan, D. Chandler,  S. Garde, \emph{J. Phys. Chem. B} {\bf 116}, 2498--2503 (2012). 
\bibitem{PDB} F.C. Bernstein, T.F. Koetzle, G.J. Williams, E.E. Meyer Jr., M.D. Brice, J.R. Rodgers, O. Kennard, T. Shimanouchi, M. Tasumi, \emph{J. of. Mol. Biol.} {\bf 112}, 535 (1977).
\bibitem{bakan} A. Bakan, L.M. Meireles,  I. Bahar, \emph{Bioinformatics} {\bf  27}, 1575 (2011).
\bibitem{cavagna2010}A. Cavagna, A. Cimarelli, I. Giardina, G. Parisi, R. Santagati, F. Stefanini, and M. Viale, \emph{PNAS} {\bf 107}, 11865 (2010).
\bibitem{attanasi} A. Attanasi, A. Cavagna, L. Del Castello, I. Giardina, S. Melillo, L. Parisi, O. Pohl, B. Rossaro, E. Shen, E. Silvestri, et al., \emph{Phys. Rev. Lett.} {\bf 113}, 238102 (2014).
 
 
\end{thebibliography}
\end{document}